\documentclass[aps,prl,twocolumn]{revtex4-2}
\usepackage{amssymb, amsmath,  hyperref, graphicx, bm, color}
\begin{document}


\title{   
High Energy Evolution of Dipole Gluon Distribution Beyond Eikonal Approximation}


\author{Ming~Li}
\affiliation{Department of Physics, Hampton University, Hampton, Virginia 23669,  USA}

\begin{abstract} 
At high energy, the dipole gluon distribution is described at eikonal order by the Wilson-line dipole correlator, whose high-energy evolution is governed by the Balitsky-Kovchegov equation. Going beyond the eikonal approximation, we identify the subeikonal operator representing the dipole gluon distribution, consisting of a Wilson line with a single insertion of the light-cone chromoelectric field, and derive its high-energy evolution equation in the large-$N_c$ limit under the single-logarithmic approximation. The resulting nonlinear evolution is coupled to the Wilson-line dipole correlator and  incorporates gluon saturation effects. In the dilute limit, the subeikonal distribution grows as a power-law of the energy with exponent $\alpha_s N_c/2\pi$, less than one fifth of the eikonal value, while in the saturation regime it obeys the same Levin--Tuchin law as the eikonal correlator. These results constitute the first closed nonlinear evolution equation for the dipole gluon distribution at subeikonal order, a step toward precision small-$x$ phenomenology and toward connecting small-$x$ physics with the moderate-$x$ dynamics.
\end{abstract}
\date{\today}
\maketitle
{\textbf{ Introduction.}}  Quantum Chromodynamics (QCD) predicts that the rapid growth of the gluon density inside a hadron toward small momentum fraction $x$ is eventually tamed by nonlinear gluon recombination, leading to gluon saturation. Saturation effects have been searched for in hadronic and nuclear collisions at RHIC and LHC, but no unambiguous signal has been established \cite{Proceedings:2026xrb}. By probing significantly smaller $x$ in a clean lepton-hadron and lepton-nucleus environment, the future Electron-Ion Collider (EIC) \cite{NSAC2023LRP,AbdulKhalek:2021gbh} is expected to reach into the saturation regime.  Exploiting this reach, however, demands a corresponding increase in accuracy of the small-$x$ QCD framework.

The theoretical description of gluon saturation rests on the high-energy evolution of Wilson-line operators. At eikonal order, the dipole gluon distribution is represented by the light-like Wilson-line dipole correlator, whose rapidity evolution is governed by the Balitsky--Jalilian-Marian--Iancu--McLerran--Weigert--Leonidov--Kovner (B-JIMWLK) equation \cite{Balitsky:1995ub, Jalilian-Marian:1997ubg,Jalilian-Marian:1997jhx,Iancu:2001ad,Iancu:2000hn}. In the large-$N_c$ limit, this evolution reduces to the Balitsky--Kovchegov (BK) equation \cite{Balitsky:1995ub,Kovchegov:1999yj}, a closed nonlinear equation that resums the leading longitudinal logarithms $\alpha_s \ln(1/x)$ to all orders while incorporating saturation. Together with cross sections computed at eikonal order and at leading order in $\alpha_s$, it forms the basis of small-$x$ phenomenology \cite{Iancu:2003xm,Gelis:2010nm}.

Two expansions control the accuracy of this framework. The perturbative expansion in the strong coupling $\alpha_s$ is well advanced: the next-to-leading-order (NLO) BK equation \cite{Balitsky:2007feb}, NLO JIMWLK equation \cite{Kovner:2013ona} are known, as are NLO cross sections for a growing set of high-energy scattering processes  (see Ref.~\cite{Proceedings:2026xrb} and references therein). The eikonal expansion in inverse powers of the center-of-mass energy squared $s$ \cite{Li:2023tlw, Chirilli:2018kkw, Chirilli:2021lif, Kovchegov:2016weo, Kovchegov:2017lsr, Cougoulic:2022gbk, Altinoluk:2021lvu, Altinoluk:2022jkk}, which relaxes the infinite-energy approximation underlying the conventional formalism, is far less developed. Yet it is precisely at the EIC, with collision energies $\sqrt{s} \simeq 20 - 140~ \mathrm{GeV}$ well below the asymptotic regime, that subeikonal corrections are expected to be numerically relevant. Their study has been hindered by the absence of closed nonlinear evolution equations, analogous to the BK equation, for the subeikonal Wilson-line correlators.

In this Letter, we derive the evolution equation of the subeikonal dipole gluon distribution in the large-$N_c$ limit and to leading logarithmic accuracy. The resulting equation is closed, couples nonlinearly to the eikonal Wilson-line dipole correlator, and consistently incorporates saturation. We further obtain its asymptotic behavior in both the dilute and saturation regimes: in the dilute regime the solution grows with a power-law exponent less than one fifth of the eikonal one, while in the saturation regime it obeys the Levin--Tuchin law \cite{Levin:1999mw}, as at eikonal order. These results open the dipole gluon distribution to quantitative study beyond the eikonal approximation in the EIC kinematics.

%
{\textbf{ Eikonal expansion of dipole gluon distribution.}}
Considering a proton moving along the negative-$z$ direction with large momentum component $P^-$,
the dipole gluon distribution in proton is defined by \cite{Dominguez:2011wm}
\begin{equation}\label{eq:Gxkt_def}
\begin{split}
&G(x, k_T^2) = \frac{2}{xP^-}\int \frac{dz^+ d^2\mathbf{z}}{(2\pi)^3} e^{ixP^- z^+} e^{-i\mathbf{k} \cdot \mathbf{z}} \\
&\times \langle P| \mathrm{tr} \left[F^{-i}(0) \mathcal{U}^{[+]}[0, z] F^{-i} (z) \mathcal{U}^{[-]}[z, 0]\right] |P\rangle_{z^-=0}
\end{split}
\end{equation}
 where the field strength tensor is $F^{-i} = \partial^- A^i - \partial^i A^- +ig[A^-, A^i]$.   Here $\mathcal{U}^{[+]}$ and $\mathcal{U}^{[-]}$ are the future- and past-pointing Wilson line staples, respectively.
\begin{equation}
 \mathcal{U}^{[\pm]}[0, z]  = V_{\mathbf{0}}[0^+, \pm \infty]  V_{\mathbf{z}}[\pm \infty, z^+].
\end{equation}
and we impose the boundary condition $A^i (z^+=\pm \infty, \mathbf{z})=0$, so that the transverse gauge links drop out \cite{Cougoulic:2022gbk}. The longitudinal gauge link in the fundamental representation is
\begin{equation}
V_{\mathbf{z}} [b^+, a^+] = \mathcal{P} \mathrm{exp}\left\{-ig \int_{a^+}^{b^+} dz^+ A_a^-(z^+, \mathbf{z})t^a\right\}.
\end{equation}
In eq.~\eqref{eq:Gxkt_def}, the dependence on $x$ enters only through the Fourier phase $e^{ixP^- z^+}$, together with the $1/x$ prefactor. At the operator level this dependence is kinematic; the dynamic $x$-dependence, associated with the high-energy evolution, is generated by loop corrections in perturbative QCD. The Fourier phase can therefore be expanded around $x \to 0$, which is an expansion in eikonality: it produces a set of operators ordered by powers of $x$, corresponding to successive orders of the eikonal expansion. At each order, the dipole gluon distribution is represented by a distinct Wilson-line operator. To organize this expansion,  let
\begin{equation}\label{eq:Lidef}
\begin{split}
L^i(x, \mathbf{z})= &-g\int_{-\infty}^{+\infty} dz^+ e^{ixP^-z^+} V_{\mathbf{z}}[+\infty, z^+]\\
&\qquad \times F^{-i}(z^+, \mathbf{z}) V_{\mathbf{z}}[z^+, -\infty].
 \end{split}
\end{equation} 
and write $L^i(x, \mathbf{z}) = \sum_{m=0}^{\infty} x^m L^i_{(m)}(\mathbf{z}) $. The first two terms are
\begin{equation}
L^i_{(0)}(\mathbf{z})=i\, \partial^i V_{\mathbf{z}}
\end{equation}
and 
\begin{equation}\label{eq:Li(1)ViG[2]}
\begin{split}
L^i_{(1)}(\mathbf{z})= V^{i, G[2]}_{\mathbf{z}}  = &-iP^- g\int dz^+ z^+ V_{\mathbf{z}}[+\infty, z^+]\\
& \times  F^{-i}(z^+, \mathbf{z}) V_{\mathbf{z}}[z^+, -\infty] \\
\end{split}
\end{equation}
with $ V_{\mathbf{z}} \equiv V_{\mathbf{z}}[+\infty, -\infty]$.  The term $L^i_{(1)}(\mathbf{z})$ is the subeikonal Wilson line $V^{i, G[2]}_{\mathbf{z}}$, carrying a single insertion of the light-cone chromoelectric field $F^{-i}$.

Using this expansion, the dipole gluon distribution takes the form 
\begin{equation}
xG(x, k_T^2) =\sum_{n=0}^{\infty} x^n  G^{(n)}(k_T^2).
\end{equation}
At the eikonal order, 
\begin{equation}
G^{(0)}(k_T^2)
= \frac{4 k_T^2}{(2\pi)^3 g^2}\int_{\mathbf{x}_1, \mathbf{x}_2} e^{-i\mathbf{k}\cdot \mathbf{x}_{12}} \left\langle \mathrm{tr}\left[ V_{\mathbf{x}_1}  V^{\dagger}_{\mathbf{x}_2}\right]\right\rangle. 
\end{equation}
Here translational invariance has been used to introduce the second coordinate $(x_2^+, \mathbf{x}_2)$.  Averaging over the proton momentum eigenstate is denoted 
$\langle \ldots \rangle = \langle P | \ldots |P\rangle /(2P^- V^+)$,  with $V^+ = \int d^2\mathbf{x}_{2}dx_2^+$;  from now on we write $(x_1^+, \mathbf{x}_1)$ in place of $(z^+, \mathbf{z})$, and set $\mathbf{x}_{12}=\mathbf{x}_1-\mathbf{x}_2$,  $\int_{\mathbf{x}} \equiv \int d^2\mathbf{x}$.  
Indeed, the dipole gluon distribution at the eikonal order is represented by the Wilson-line dipole correlator
\begin{equation}
S(\mathbf{x}_1, \mathbf{x}_2) = \frac{1}{N_c}\left\langle \mathrm{tr}\left[ V_{\mathbf{x}_1}  V^{\dagger}_{\mathbf{x}_2}\right]\right\rangle.
\end{equation}
At subeikonal order, the dipole gluon distribution is represented by
\begin{equation}\label{eq:G(1)_ViG[2]}
\begin{split}
G^{(1)}(k_T^2) &= \frac{4}{(2\pi)^3 g^2} \mathbf{k}^i \int_{\mathbf{x}_1, \mathbf{x}_2}e^{-i\mathbf{k}\cdot \mathbf{x}_{12}}\\
&\times \left\{\left\langle \mathrm{tr}\left[V_{\mathbf{x}_1}V^{i\, G[2]\dagger}_{\mathbf{x}_2}\right]\right\rangle  + \left\langle \mathrm{tr}\left[V^{i\, G[2]}_{\mathbf{x}_1}V^{\dagger}_{\mathbf{x}_2}\right] \right\rangle\right\}.\\
\end{split}
\end{equation}
Under translational invariance, the chromoelectric Wilson-line dipole can be further decomposed as 
\begin{align}\label{eq:DEix12}
D_E^i(\mathbf{x}_1, \mathbf{x}_2)=&\frac{1}{N_c} \left\langle \mathrm{tr}\left[V^{i\, G[2]}_{\mathbf{x}_1}V^{\dagger}_{\mathbf{x}_2}\right] \right\rangle \\
= &\mathbf{x}_{12}^i\, D_E^{\parallel}(|\mathbf{x}_{12}|) + \epsilon^{ij} \mathbf{x}_{12}^j \, D_E^{\perp}(|\mathbf{x}_{12}|). \label{eq:DEidecomp}
\end{align}
Substituting eq.~\eqref{eq:DEidecomp}
 into eq.~\eqref{eq:G(1)_ViG[2]}, the $D_E^{\perp}$ component drops out, and only the imaginary part  $\mathrm{Im}D_{E}^{\parallel}(|\mathbf{x}_{12}|)$ contributes
   \begin{equation}
G^{(1)}(k_T^2) 
=\frac{-8N_c S_{\perp}}{(2\pi)^3 g^2} \,  \mathbf{k}^i \frac{\partial}{\partial \mathbf{k}^i}  \mathrm{Im} D_E^{\parallel} (k_T^2)
\end{equation}
where $S_{\perp}$ is the transverse area. 
Eq.~\eqref{eq:DEix12} is the operator representing the dipole gluon distribution at subeikonal order. As with the Wilson-line dipole correlator at eikonal order, the aim of this Letter is to derive the nonlinear small-$x$ evolution equation obeyed by this operator and to determine its  small-$x$ asymptotic behavior.

%
%
 
 {\textbf{High energy evolution equation of subeikonal dipole gluon distribution}.} To derive the small-$x$ evolution equation for the subeikonal Wilson-line correlator in eq.~\eqref{eq:DEix12}, we use the background-field method in the nuclear shockwave formalism \cite{Balitsky:1995ub, Kovchegov:2017lsr}, which is essentially the Wilsonian approach to renormalization-group equation \cite{Peskin:1995ev}. It is convenient to rewrite eq.~\eqref{eq:Li(1)ViG[2]} as
 \begin{equation}
V_{\mathbf{z}}^{i, G[2]}  
=igP^- \int dz^+  V_{\mathbf{z}}[+\infty, z^+] \widetilde{A}^i(z^+, \mathbf{z}) V_{\mathbf{z}}[z^+, -\infty],
\end{equation}
introducing the subeikonal-order effective field $\widetilde{A}^i = z^{+} \partial^i A^-+ A^i$.  Unlike the eikonal Wilson-line dipole, whose evolution involves only the eikonal field $A^-$, the subeikonal operator $O_E^i$ in the definition $D_E^i = \langle O^i_E \rangle$ depends on both  $A^-$ and $\widetilde{A}^i$. 

At rapidity $Y$, the operator $O_E^i$ contains gluon modes with longitudinal momentum up to $p^+_{\mathrm{max}}$. Increasing the rapidity by a small interval $\Delta Y$ introduces additional gluon modes $A^{u} \rightarrow A^{u}+\mathcal{A}^{u}$.  We denote $A^{u}\equiv (A^-, \widetilde{A}^i)$. Here $\mathcal{A}^{u}\equiv (\mathcal{A}^-, \widetilde{\mathcal{A}}^i)$ represents the gluon modes whose longitudinal momentum lie in the small rapidity strip $[p^+_{\mathrm{max}}, e^{\Delta Y} p^+_{\mathrm{max}}]$.  The evolution equation is obtained by integrating out the gluon modes in this rapidity strip. Formally, at the operator level, the evolution is generated by
\begin{equation*}
\begin{split}
&O^i_E[ A^{u}+\mathcal{A}^{u}]_{Y+ \Delta Y}= O^i_E[A^{u}]_{Y}  +\frac{\delta^2 O_E^i}{\delta A^u\delta A^v} \langle \mathcal{A}^u \mathcal{A}^v\rangle_{\mathcal{B}} + \ldots
\end{split}
\end{equation*}
The tadpole diagram vanishes, $\langle \mathcal{A}^{u}\rangle_{\mathcal{B}}=0$.  The evolution  therefore reduces to evaluating the background-field propagator $\langle A^{u} A^{v} \rangle_{\mathcal{B}}$. Averaging $ O_E^i$ over the proton wavefunction then gives the evolution of $D_E^i$.
 
\begin{figure}
    \centering
    \includegraphics[width=0.45\textwidth]{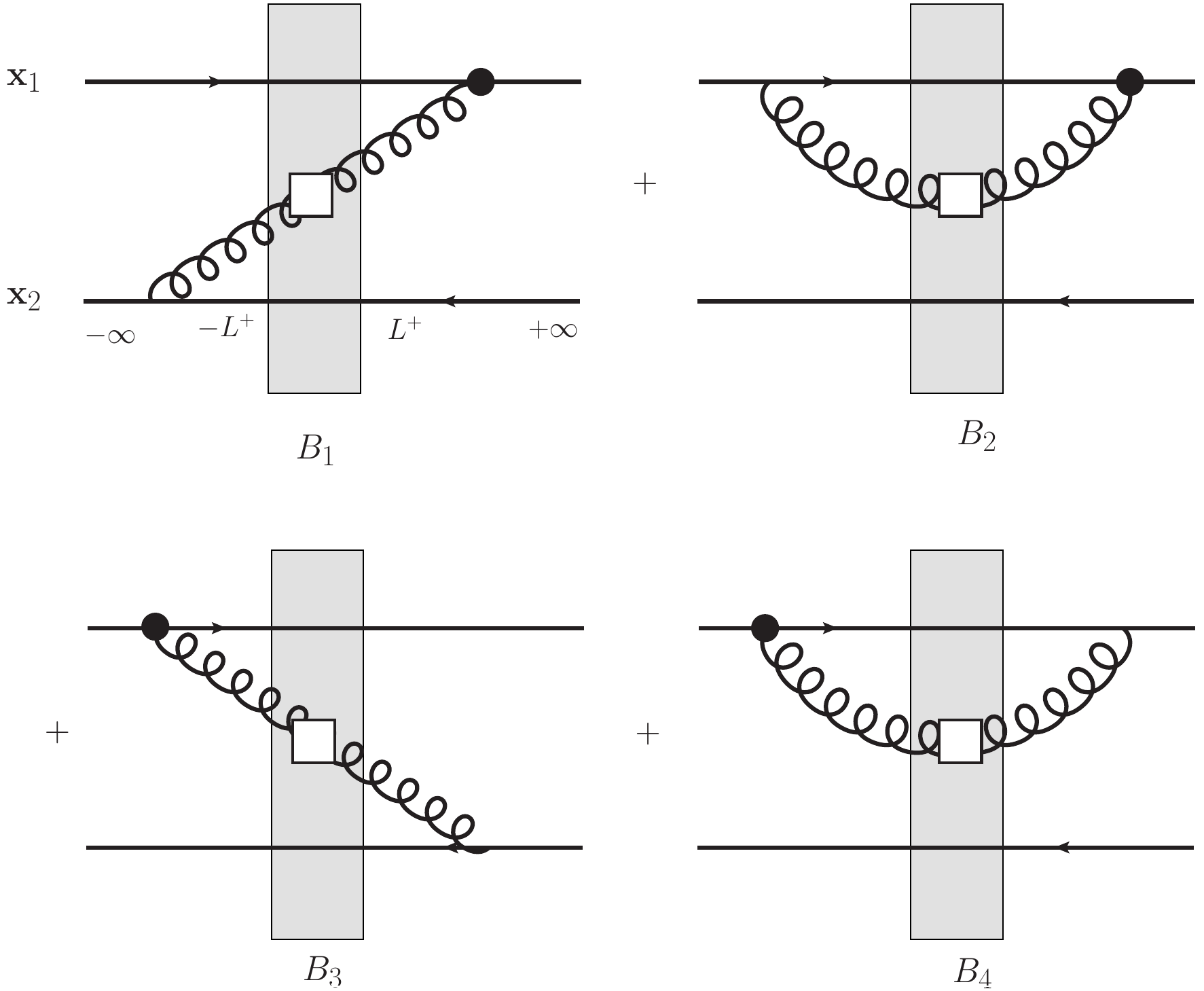}
    \caption{Real-emission diagrams contributing to the evolution of the chromoelectric Wilson-line dipole correlator. The gray rectangle denotes the shockwave background, while the white square represents the subeikonal interactions.  The solid black dot indicates subeikonal order gluon emission vertex. }
	\label{fig:dipole_subeikonal}
\end{figure}

 Since $O_E^i$ is linear in $\widetilde{A}^i$, the propagator $\langle \widetilde{\mathcal{A}}^i \widetilde{\mathcal{A}}^j\rangle_{\mathcal{B}}$ does not contribute. The eikonal propagator $\langle \mathcal{A}^- \mathcal{A}^-\rangle_{\mathcal{B}}$ is the same propagator that generates the BK equation.
 Its contribution to the subeikonal evolution has exactly the same structure as in the BK equation, with the eikonal Wilson line $V_{\mathbf{x}_1}$ replaced by its subeikonal counterpart $V^{i, G[2]}_{\mathbf{x}_1}$. The resulting contribution reads  
 \begin{equation}\label{eq:BK-type}
\begin{split}
&\Big\langle\frac{\delta^2 O_E^i}{\delta A^-\delta A^-} \langle \mathcal{A}^- \mathcal{A}^-\rangle_{\mathcal{B}}\Big\rangle =   \frac{\alpha_s}{\pi^2} \int_{p_1^+, \mathbf{x}_0} \frac{|\mathbf{x}_{12}|^2}{|\mathbf{x}_{10}|^2|\mathbf{x}_{20}|^2} \\
&\times \left\{\left\langle U^{ab}_{\mathbf{x}_0} \mathrm{tr}\left[V^{i, G[2]}_{\mathbf{x}_1} t^b V^{\dagger}_{\mathbf{x}_2} t^a\right] \right\rangle -C_F \left\langle\mathrm{tr}\left[V^{i, G[2]}_{\mathbf{x}_1}V^{\dagger}_{\mathbf{x}_2}\right] \right\rangle \right\}
\end{split}
\end{equation}
where $C_F = (N_c^2-1)/2N_c$ and $\int_{p_1^+} \equiv \int dp_1^+/p_1^+$.  The genuine new contributions come from the subeikonal propagator $\langle \mathcal{A}^- \widetilde{\mathcal{A}}^i\rangle_{\mathcal{B}}$, which generates the four real-emission diagrams shown in Fig.~\ref{fig:dipole_subeikonal}.
As an illustration, the propagator corresponding to diagram-$B_1$ reads
\begin{equation*}
\begin{split}
&g^2 P^-\int_{L^+}^{+\infty} dx_1^+\int^{-L^+}_{-\infty} dx_2^+\left\langle \widetilde{\mathcal{A}}_a^j(x_1^+, \mathbf{x}_1)\mathcal{A}^-_b(x_2^+, \mathbf{x}_2)\right\rangle_{\mathcal{B}}\\
=& \frac{-\alpha_s}{4\pi^2}\int_{p_1^+,\mathbf{x}_0} h^{jl}(\mathbf{x}_{10})\Big\{  f^l(\mathbf{x}_{20})  \left( is U_{\mathbf{x}_0}  + iU_{\mathbf{x}_0}^{q[2]} -U^{G[3]}_{\mathbf{x}_0} \right)^{ab} \\
& - \epsilon^{kl} f^k(\mathbf{x}_{20}) U^{\mathrm{pol} [1], ab}_{\mathbf{x}_0}  + i \int_{ \mathbf{x}_{0'}}  f^l(\mathbf{x}_{20'}) U^{G[2],ab}_{\mathbf{x}_0, \mathbf{x}_{0'}} \Big\}.
\end{split}
\end{equation*}
 Here $f^l (\mathbf{x}) = \mathbf{x}^l/|\mathbf{x}|^2$ and $h^{jl}(\mathbf{x}) = \delta^{jl} - 2\mathbf{x}^j\mathbf{x}^l/|\mathbf{x}|^2$. $s = 2p_1^+ P^-$ is the center-of-mass collision energy squared.  The dependence on $L^+$ drops out after imposing the kinematic constraint $\mathbf{p}_1^2/2p_1^+ \gg 1/L^+ \sim P^-$ and concentrating on single logarithmic terms. The integrations over $p_1^+, \mathbf{x}_0$ are not independent and must satisfy the kinematic constraint \cite{Beuf:2014uia}. The subeikonal order Wilson lines $U^{\mathrm{pol}[1]}_{\mathbf{x}_0} = U^{q[1]}_{\mathbf{x}_0} + U^{G[1]}_{\mathbf{x}_0}$, $U^{q[2]}_{\mathbf{x}_0}$, $U^{G[2]}_{\mathbf{x}_0, \mathbf{x}_{0'}}$, $U^{G[3]}_{\mathbf{x}_0}$ have been extensively studied in various forms in \cite{Li:2023tlw, Chirilli:2018kkw, Chirilli:2021lif, Kovchegov:2016weo, Kovchegov:2017lsr, Cougoulic:2022gbk, Altinoluk:2021lvu, Altinoluk:2022jkk}. Their explicit expressions can be found in \cite{Li:2023tlw, Cougoulic:2022gbk, Kovchegov:2025gcg}. 
Summing the four real-emission diagrams in Fig.~\ref{fig:dipole_subeikonal} yields
 \begin{equation}\label{eq:result_real}
\begin{split}
&\Big\langle\frac{\delta^2 O_E^i}{\delta \widetilde{A}^j\delta A^-} \langle \mathcal{\widetilde{A}}^j \mathcal{A}^-\rangle_{\mathcal{B}}\Big\rangle= \frac{\alpha_s}{4\pi^2}\int_{p_1^+,\mathbf{x}_0}h^{il}(\mathbf{x}_{10})\Big\{ 2g^k(\mathbf{x}_{10}, \mathbf{x}_{20})\\
&\qquad  \times \Big\langle \left( \epsilon^{kl} U^{\mathrm{pol} [1]}_{\mathbf{x}_0} + \delta^{kl} U^{G[3]}_{\mathbf{x}_0}\right)^{ab}\mathrm{tr}[t^aV_{\mathbf{x}_1} t^bV_{\mathbf{x}_2}^{\dagger}]\Big\rangle \\
&-i \int_{\mathbf{x}_{0'}}  g^l(\mathbf{x}_{10'}, \mathbf{x}_{20'})\Big\langle \left(U^{G[2]}_{\mathbf{x}_0, \mathbf{x}_{0'}} -U_{\mathbf{x}_{0'}, \mathbf{x}_0}^{G[2]}\right)^{ab}\mathrm{tr}[t^aV_{\mathbf{x}_1} t^bV_{\mathbf{x}_2}^{\dagger}]\Big\rangle\Big\}.
\end{split}
\end{equation}
Here $g^{k}(\mathbf{x}_{10}, \mathbf{x}_{20}) = f^k(\mathbf{x}_{10}) - f^k(\mathbf{x}_{20})$. Eq.~\eqref{eq:result_real} contains all genuinely new operator structures. Notice that terms containing the eikonal Wilson line $U_{\mathbf{x}_0}$ and the subeikonal Wilson line $U^{q[2]}_{\mathbf{x}_0}$ cancel out after adding the four diagrams. The terms involving $U^{G[2]}$ can be further simplified using the identity
\begin{equation*}
\begin{split}
&\int_{\mathbf{x}_{0'}, \mathbf{x}_{0}}  g^l(\mathbf{x}_{10'}, \mathbf{x}_{20'}) h^{il}(\mathbf{x}_{10}) \left(U^{G[2]}_{\mathbf{x}_0, \mathbf{x}_{0'}} -U_{\mathbf{x}_{0'}, \mathbf{x}_0}^{G[2]}\right)\\
=&2i\int_{\mathbf{x}_0}  \left[g^l(\mathbf{x}_{10}, \mathbf{x}_{20}) (\overleftarrow{\partial}^j_{\mathbf{x}_0} - \overrightarrow{\partial}^j_{\mathbf{x}_0}) h^{il}(\mathbf{x}_{10})\right] U^{j, G[2]}_{\mathbf{x}_0}.
\end{split}
\end{equation*}
As shown below, most of these operators disappear after taking the large-$N_c$ limit and projecting onto $\mathrm{Im}D_{E}^{\parallel}$. 
 Combining eqs.~\eqref{eq:BK-type} and ~\eqref{eq:result_real} yields the evolution equation for $D_E^i$. Related partial results were obtained previously in the derivation of the gluon helicity evolution equation \cite{Cougoulic:2022gbk}. 
 
 As with BK equation, the evolution does not close at finite $N_c$, because the subeikonal propagator generates additional subeikonal Wilson-line operators beyond $D_E^i$. We therefore resort to taking the large-$N_c$ limit. Our primary interest, however, is the evolution equation for $\mathrm{Im} D_{E}^{\parallel}(x_{12})$ (denoting $x_{12} \equiv |\mathbf{x}_{12}|$)
\begin{equation*}
\mathrm{Im} D_{E}^{\parallel}(x_{12}) = \frac{1}{2i}\frac{\mathbf{x}_{12}^i}{|\mathbf{x}_{12}|^2}\left[D_E^i(\mathbf{x}_1, \mathbf{x}_2) + D_E^{i\ast}(\mathbf{x}_2, \mathbf{x}_1)\right].
\end{equation*}
 In the large-$N_c$ limit, the adjoint-representation eikonal and subeikonal Wilson lines are expressed through fundamental-representation Wilson lines, and the evolution of $\mathrm{Im}D_E^{\parallel}(x_{12})$ simplifies substantially: the terms involving $U^{\mathrm{pol}[1]}_{\mathbf{x}_0}$ and $U^{G[3]}_{\mathbf{x}_0}$ vanish because of the cross-product appearing in the kernel. For instance,
 \begin{equation}
 \int_{\mathbf{x}_0} \frac{\mathbf{x}_{10}\times \mathbf{x}_{20}}{|\mathbf{x}_{10}|^2 |\mathbf{x}_{20}|^2} \left\langle \mathrm{tr}[V_{\mathbf{x}_1}V_{\mathbf{x}_0}^{\dagger}]\right\rangle \left\langle \mathrm{tr}[V_{\mathbf{x}_0}^{G[3]}V_{\mathbf{x}_2}^{\dagger}]\right\rangle = 0
 \end{equation}
 The vanishing follows from translational and rotational invariance on the transverse plane, under which the eikonal and subeikonal correlators depend only on the corresponding dipole sizes.  The same argument eliminates the contributions involving $V^{\mathrm{pol}[1]}_{\mathbf{x}_0}$.  Only $V_{\mathbf{x}}^{i, G[2]}$ then survives, and the evolution closes. 
  
 Carrying out the longitudinal-momentum integral, $\int dp_1^+/p_1^+ =\Delta Y$,  and taking the limit $\Delta Y\rightarrow 0$ gives the differential evolution equation (with $\mathcal{D} \equiv \mathrm{Im}D_E^{\parallel}$) 
\begin{equation}\label{eq:main_result}
\begin{split}
&\frac{\partial\mathcal{D}(x_{12}, Y) }{\partial Y}  
= \frac{\alpha_s N_c }{2\pi^2}  \int_{\mathbf{x}_0}  \mathcal{K}_1(\mathbf{x}_{10}, \mathbf{x}_{20})S(x_{10}, Y)\, \mathcal{D}(x_{20}, Y)\\
&+ \mathcal{K}_1(\mathbf{x}_{20}, \mathbf{x}_{10}) S(x_{20}, Y)\, \mathcal{D}(x_{10}, Y) -  \mathcal{K}_0(\mathbf{x}_{10}, \mathbf{x}_{20})\mathcal{D}(x_{12}, Y), \\
\end{split}
\end{equation}
and the kernels 
\begin{align}
&\mathcal{K}_0(\mathbf{x}_{10}, \mathbf{x}_{20})= \frac{|\mathbf{x}_{12}|^2}{|\mathbf{x}_{10}|^2|\mathbf{x}_{20}|^2},\\
&\mathcal{K}_1(\mathbf{x}_{10}, \mathbf{x}_{20}) = -\frac{\mathbf{x}_{12}\cdot \mathbf{x}_{20}}{2|\mathbf{x}_{10}|^2|\mathbf{x}_{20}|^2}+ \frac{(\mathbf{x}_{10}\times \mathbf{x}_{20})^2}{|\mathbf{x}_{10}|^4|\mathbf{x}_{20}|^2}.
\end{align}
To confirm that Eq.~\eqref{eq:main_result} is genuinely single-logarithmic and takes this differential form, one must check that no further transverse logarithms $\int d r^2_{\perp}/r_{\perp}^2$ appear on the right-hand side.  Such transverse integrals would, after imposing the kinematic constraint, generate double logarithms of energy \cite{Chirilli:2026pkv} and spoil the single-logarithmic approximation. Following the standard analysis of the potentially logarithmic regions, it is straightforward to verify that no such transverse integrals survive either in the limits $\mathbf{x}_0 \rightarrow \mathbf{x}_1, \mathbf{x}_{0}\rightarrow \mathbf{x}_2$,  where one daughter dipole becomes much smaller than the parent dipole or in the region $\mathbf{x}_{10}\sim \mathbf{x}_{20} \gg \mathbf{x}_{12}$ where both daughter dipoles are much larger than the parent dipole . Equation \eqref{eq:main_result} therefore describes the leading single-logarithmic evolution of the subeikonal dipole gluon distribution.
 
 Eq.~\eqref{eq:main_result} constitutes the central result of this paper.  In the large-$N_c$ limit,  the evolution equation is closed once supplemented by the BK equation for the Wilson-line dipole correlator $S(x_{\perp}, Y)$. The equation is linear in $\mathcal{D}(x_{\perp}, Y)$, but  gluon saturation effects enter through its coupling to $S(x_{\perp},Y)$. The structure of eq.~\eqref{eq:main_result} is reminiscent of the high-energy evolution equation for the odderon \cite{Kovchegov:2003dm,Hatta:2005as};  however, the latter involves only the eikonal kernel $\mathcal{K}_0$, whereas eq.~\eqref{eq:main_result} contains the new kernel $\mathcal{K}_1$.

%
%
\begin{figure}
    \centering
    \includegraphics[width=0.45\textwidth]{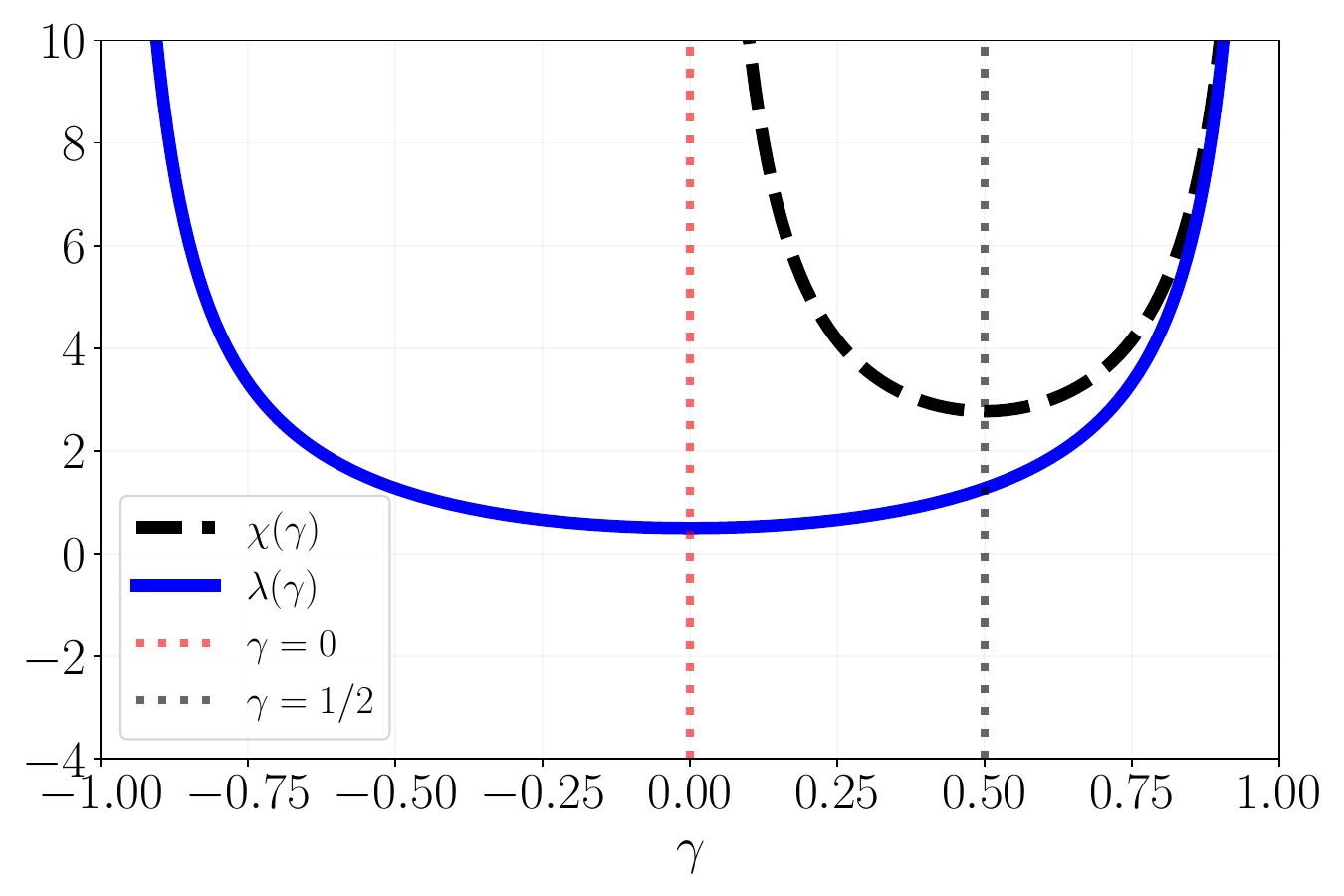}
    \caption{ Comparison of the eigenvalue $\lambda(\gamma)$ governing the subeikonal evolution in the range $-1<\gamma<1$ with the BFKL eigenvalue $\chi(\gamma)$ in the range $0<\gamma<1$. Unlike $\chi(\gamma)$, $\lambda(\gamma)$ is finite at $\gamma=0$. }
	\label{fig:lambda_chi}
\end{figure}

{\textbf{ Asymptotic solutions}.} 
For phenomenological applications, eq.~\eqref{eq:main_result} must be solved together with the BK equation, but its large-rapidity asymptotics can be determined semi-analytically, following the standard analysis of BK solutions reviewed in \cite{Kovchegov:2012mbw}. In the dilute regime, $x_{\perp}^2 Q_{s0}^2\ll 1$ with $Q_{s0}$ representing the non-perturbative infrared scale,  one has $S(x_{10}, Y) \sim S(x_{20}, Y) \sim 1$, and eq.~\eqref{eq:main_result} becomes linear. Since the evolution kernels are scale invariant, the evolution equation is diagonalized by the Mellin eigenfunctions. Performing the Mellin transform
\begin{equation}
\mathcal{D}(x_{\perp}, Y) = \int_{a-i\infty}^{a+i\infty} \frac{d\gamma}{2\pi i} (x_{\perp}^2)^{\gamma} \widetilde{\mathcal{D}}(\gamma, Y),
\end{equation}
The linearized evolution equation becomes
\begin{equation}
\frac{\partial}{\partial Y} \widetilde{\mathcal{D}}(\gamma, Y) = \bar{\alpha}_s \, \lambda(\gamma) \widetilde{\mathcal{D}}(\gamma, Y)
\end{equation}
where $\bar{\alpha}_s = \alpha_s N_c/ \pi$ and $\lambda(\gamma)$ is fixed by
\begin{equation}
\begin{split}
&\frac{1}{2\pi} \int_{\mathbf{x}_0} \Big\{\mathcal{K}_1(\mathbf{x}_{10}, \mathbf{x}_{20}) \, (x_{20}^2)^{\gamma} + \mathcal{K}_1(\mathbf{x}_{20}, \mathbf{x}_{10}) \, (x_{10}^2)^{\gamma} \\
&\qquad\quad- \mathcal{K}_0(\mathbf{x}_{10}, \mathbf{x}_{20})\, (x_{12}^2)^{\gamma}\Big\}=\lambda(\gamma) \, (x_{12}^2)^{\gamma} .
\end{split}
\end{equation}
Using the standard Mellin techniques reviewed in \cite{Kovchegov:2012mbw}, one obtains
\begin{equation}
\begin{split}
\lambda(\gamma)  =& 2\psi(1) - \psi(1+\gamma) - \psi(1-\gamma) + \frac{1}{2}\\
=&\chi(\gamma) - \frac{1}{\gamma} + \frac{1}{2}
\end{split}
\end{equation}
where $\chi(\gamma) = 2\psi(1) - \psi(\gamma) -\psi(1-\gamma)$ is the BFKL eigenvalue with $\psi(\gamma)$ denoting the digamma function. The $1/\gamma$ pole cancels, so $\lambda(\gamma)$ is finite at $\gamma=0$ with $\lambda(0) = 1/2$. In BFKL evolution the intercept comes from the saddle point at $\gamma=1/2$, giving $4\ln 2$; here the small-$x$ behavior is instead controlled by the finite value at $\gamma=0$, which is the origin of the much smaller exponent found below. See the illustrative comparison of $\lambda(\gamma)$ and $\chi(\gamma)$ in fig.~\ref{fig:lambda_chi}. 
 
Unlike the BFKL problem, the Mellin contour is initially determined by the Born-level behavior of the subeikonal operator. A straightforward Born-level calculation gives that $\mathcal{D}(x_{\perp}, Y=0) \sim \ln (1/x_{\perp}^2 Q_{s0}^2)$ when $x_{\perp}\rightarrow 0$. This requires that $\mathrm{Re} (\gamma)<0$ and $\widetilde{\mathcal{D}}(\gamma, Y=0) = C_{\gamma}/\gamma^2$ contains a double pole. Here $C_{\gamma} = C_0 + C_1 \gamma + C_2\gamma^2+ \ldots$ is analytic in $\gamma$. The Mellin contour is then analytically continued from $-1<\mathrm{Re}(\gamma)=a<0$ to $0<\mathrm{Re}(\gamma)=c<1$. In doing so, the contour crosses the double pole in the initial condition whose residue must be included explicitly.  The Mellin solution becomes
\begin{equation}\label{eq:sol_Mellin_3}
\mathcal{D}(\rho, \eta) = e^{\eta/2}\Big(C_0\rho - C_1\Big)+  \int_{c-i\infty}^{c+i\infty} \frac{d\gamma}{2\pi i} \frac{C_{\gamma} }{\gamma^2} e^{  \lambda(\gamma)\eta - \gamma \rho }
\end{equation}
with $\eta =\bar{\alpha}_s Y$ and $\rho =  \ln (1/x_{\perp}^2 Q_{s0}^2)$.  To extract the large $Y$ asymptotic behavior, we consider $\eta \gg \rho$, in which case the saddle point $\gamma_{\ast}$ obtained by setting $\lambda'(\gamma_{\ast}) = \rho/\eta$ approaches the initial-condition pole $\gamma=0$. The usual saddle-point approximation breaks down. The Mellin integral is therefore evaluated using the uniform asymptotic expansion for saddle-pole coalescence \cite{Bleistein1966, Wong2001Asymptotic}: expanding the phase factor around $\gamma=0$ to quadratic order and computing the resulting integrals in terms of the error function. For $\eta\gg \rho$, 
\begin{equation}
\mathcal{D}( \rho, \eta) 
\simeq e^{\eta/2} \Big[ C_0 \sqrt{\frac{2\zeta(3)\eta}{\pi}}  +\frac{C_0 \rho - C_1}{2} + \frac{C_2}{\sqrt{2\pi} 4\zeta(3) \eta} + \ldots \Big].
\end{equation}
The leading asymptotic behavior at large rapidity is therefore set by the exponent $\bar{\alpha}_s/2$, which is less than one fifth of the eikonal exponent $4\ln 2 \, \bar{\alpha}_s$.

For large dipoles in the saturation regime, $S(x_{10}, Y) \sim S(x_{20}, Y) \sim 0$, the first two terms of eq.~\eqref{eq:main_result} are suppressed and the $-\mathcal{K}_0(\mathbf{x}_{10}, \mathbf{x}_{20}) \, \mathcal{D}(x_{12})$ term  dominates. The evolution equation then reduces to the same form as  that of the eikonal BK equation. The saturation regime is therefore governed by the Levin--Tuchin law \cite{Levin:1999mw}, as at eikonal order, even though the dilute regime carries a new growth exponent.

%

{\textbf{Conclusions}.}  
In this Letter, we identified the subeikonal operator representing the dipole gluon distribution and derived its high-energy evolution equation. Together with the BK equation for the eikonal Wilson-line dipole correlator, the new evolution equation forms a closed system that consistently incorporates gluon saturation effects. We further determined  the asymptotic behavior of the solution analytically in both the dilute and saturation regimes. Quantitative applications require solving the coupled evolution equations numerically with appropriate initial conditions, which we leave to future work. 

An important open question is how the subeikonal dipole gluon distribution can be accessed experimentally. At eikonal order, the inclusive deep-inelastic scattering (DIS) cross section directly probes the dipole gluon distribution. Recent studies indicate that this direct correspondence no longer holds at subeikonal order \cite{Altinoluk:2025ivn, Chirilli:2026vij}. More differential observables, involving particle production in DIS and in hadron-hadron collisions, are expected to be sensitive to the subeikonal dipole gluon distribution \cite{Altinoluk:2024dba, Kovchegov:2024aus}.

 More generally, subeikonal interactions generate a larger set of operators than at eikonal order, including genuinely subeikonal operators for the quark and antiquark distributions at small $x$. Their high-energy evolution generally involves operator mixing, requiring a coupled system of evolution equations \cite{Borden:2024bxa,Chirilli:2021lif}. Establishing this broader set of evolution equations will be necessary for quantitative phenomenology beyond the eikonal approximation, and the equation derived in this Letter is a first step toward such a framework.

The eikonal expansion is equivalent to an expansion in powers of $x$, so incorporating higher-order corrections extends the reach of the small-$x$ framework toward moderate $x$. The resummation achieved through the subeikonal evolution equation deepens the connection with the twist expansion of collinear factorization \cite{Fu:2023jqv}, and helps bridge small-$x$ physics with the transverse-momentum-dependent description of hadron structure \cite{Balitsky:2015qba,Balitsky:2016dgz,Mukherjee:2023snp}.


{\bf Acknowledgments.}
 The author is grateful to Yuri Kovchegov for stimulating discussions. The author thanks the Theory Center at Jefferson Lab for its hospitality, where part of this work was completed.  


  \bibliography{softgluon}
\end{document}